\documentclass{revtex4}
\usepackage{amssymb}
\usepackage{amsmath}

\input{tcilatex}

\begin{document}

\title{Effect of the Refractive Index on the Hawking Temperature: An
Application of the Hamilton-Jacobi Method }
\author{I. Sakalli$^{\dag }$ and S.F. Mirekhtiary$^{\ast }$}
\affiliation{Department of Physics, Eastern Mediterranean University,}
\affiliation{G. Magusa, North Cyprus, Mersin-10, Turkey}
\affiliation{$^{\dag }$izzet.sakalli@emu.edu.tr,$^{\ast }$fatemeh.mirekhtiary@emu.edu.tr}

\begin{abstract}
Hawking radiation of a non-asymptotically flat (NAF) 4-dimensional
spherically symmetric and static dilatonic black hole (BH) via the
Hamilton-Jacobi (HJ)\ method has been studied. In addition to the naive
coordinates, we have used four more different coordinate systems which are
well-behaved at the horizon. Except the isotropic coordinates, direct
computation of the HJ method leads us the standard Hawking temperature for
all coordinate systems. The isotropic coordinates render possible to get the
index of refraction extracting from the Fermat metric. It is explicitly
shown that the index of refraction determines the value of the tunneling
rate and its natural consequence, Hawking temperature. The isotropic
coordinates within the conventional HJ method produce wrong result for the
temperature of the dilatonic BH. Here, we explain how this discrepancy can
be resolved by regularizing the integral possessing a pole at the horizon.
\end{abstract}

\maketitle

\section{\textbf{Introduction}}

In 1974, Stephan Hawking \cite{Hawking,Hawking2} proved that a BH, when
taking into account of quantum effects, can emit thermal radiation. This
meant that each BH has a characteristic temperature and can be thought as a
thermodynamical system. In fact, this discovery broke\ all taboos which were
classically prohibited about the BHs until that day. Together with
Bekenstein's work \cite{Bekenstein} it caused to born a new subject that is
the so-called quantum gravity theory which has not been completed yet. After
Hawking, there has always been curiosity to derive new methods to the
Hawking radiation (HR) which can decode\ the underlying BH\ spacetime.
Today, we can see many found methods for the HR in the literature (see \cite%
{DonPage} and references therein for a general review). Among them, the most
promising one is the tunneling method which is originated from Kraus and
Wilczek (KW) \cite{KW1,KW2}. KW used the null geodesic method to develop the
action for the tunneling particle which is considered as a self-gravitating
thin spherical shell and then managed to quantize it. KW method's strong
suit indeed is to provide a dynamical model of the HR in which BH shrinks as
particles radiate. In this dynamical model, both energy conservation and
back-reaction effects are included which were not considered in the original
derivation of HR. Six years later, their calculations were reinterpreted by
Parikh and Wilczek (PW) \cite{PW}. They showed that the spectrum of the HR
can deviate from pure thermality, which implies unitarity of the underlying
quantum process and the resolution of the information loss paradox \cite%
{ILP1,ILP2}. Nowadays, PW's pioneer work has been still preserving its
popularity. A lot of works for various BHs proves its validity (a reader may
refer to \cite{Vanzo}). As far as we know, the original PW's tunneling
method only suffers from one of the NAF BHs which is the so-called linear
dilaton BH (LDBH). Unlike to the other well-known BHs, their evaporation
does not admit non-thermal radiation, therefore causes the violation of
information conservation. This problem was firstly unraveled by Pasaoglu and
Sakalli \cite{Pasaoglu}. Recently, it has been shown that the weakness of
the PW's method in retrieving the information from the LDBH can be overcome
by adding the quantum corrections to the entropy \cite{Sakalli1}.
Furthermore, it has proven by another study of Sakalli et al. \cite{Sakalli2}
that the entropy of the LDBH can be tweaked by the quantum effects that both
its temperature and mass simultaneously become zero at the end of the
complete evaporation.

Based on the complex path analysis of Padmanabhan and his collaborators \cite%
{Padnab1,Padnab2,Padnab3}, Angheben \ et al. \cite{Angheben} developed an
alternate method for calculating the imaginary part of the action belonging
to the tunneling particles. To this end, they made use of the relativistic
HJ equation. Their method neglects the effects of the particle
self-gravitation and involves the WKB approximation. In general, the
relativistic HJ equation can be solved by substituting a suitable ansatz.
The chosen ansatz should consult the symmetries of the spacetime in order to
allow for the separability. Thus one can get a resulting equation which is
solved by integrating along the classically forbidden trajectory that
initiates inside the BH and ends up at the outside observer. However, the
integral has always a pole located at the horizon. For this reason, one
needs to apply the method of complex path analysis in order to circumvent
the pole.

A Friedmann-Robertson-Walker universe --- assumed to be a good model for our
universe --- is generally NAF \cite{FRW}. For this reason, we believe that
most of the BHs in the real universe have necessarily NAF geometries. Hence,
it is of our special interest to find out specific examples of NAF BHs as a
test bed for HR\ problems via the HJ method. Starting from this idea, in
this paper we consider the LDBHs. First of all, the eponym of these BHs is Cl%
\'{e}ment and Gal'tsov \cite{Clement}. Initially, they were found as a
solution to Einstein-Maxwell-Dilaton (EMD) theory \cite{Chan} in four
dimensions. Later on, it is shown that in addition to the EMD theory $N\geq
4 $ dimensional LDBHs (even in the case of higher dimensions) are available
in Einstein-Yang-Mills-Dilaton (EYMD) and
Einstein-Yang-Mills-Born-Infeld-Dilaton (EYMBID) theories \cite{MSH} (and
references therein). The most intriguing feature of these BHs is that while
radiating, they undergo an isothermal process. Namely, their temperature
does not alter with shrinking of the BH horizon or with the mass loss. Our
primary concern of this study is to obtain the imaginary part of the action
of the tunneling particle through the LDBH's horizon. This produces the
tunneling rate that yields the Hawking temperature. In order to test the HJ
method on the LDBH, in addition to the naive coordinates we will consider
four more coordinate systems (all regular) which are isotropic, Painlev\'{e}%
-Gullstrand (PG), ingoing Eddington-Finkelstein (IEF) and Kruskal-Szekeres
(KS), respectively. Especially, we will mainly focus on the isotropic
coordinates. They require more straightforward calculations when compared
with the others. Furthermore, as it will be shown in the associated section,
the usage of the standard HJ method with isotropic coordinates reveals a
discrepancy in the temperatures. For a more recent account in the same line
of thought applied to Schwarzschild BH within the isotropic coordinates, one
may consult \cite{GRG12} in which a similar discrepancy problem in the HR
has been studied. Gaining inspiration from \cite{GRG12}, we will discuss
about how one can also remove the discrepancy appeared in the LDBH's
radiation. Differently from \cite{GRG12}, we will also represent the
calculation of the index of refraction of the LDBH medium, its effect on the
tunneling rate and consequently on the Hawking temperature. According to our
knowledge, such a theoretical observation has not been reported before in
the literature. Slightly different from the other coordinate systems, during
the application of the HJ method in the KS coordinates, we will first reduce
the LDBH spacetime to Minkowski space and then demonstrate in detail how one
recovers the Hawking temperature.

The paper uses the signature $(-,+,+,+)$ and units where $c=G=\hbar =k_{B}=1$%
. The paper is organized as follows. In Sec. II we review some of the
geometrical and thermodynamical features of the LDBH with naive coordinates
and show the separation of variables of the relativistic HJ equation. The
calculation of the tunneling rate and henceforth the Hawking temperature via
the HJ method is also represented. In Sec. III the metric for a LDBH in
isotropic coordinates is derived. The effect of index of refraction on the
tunneling rate is explicitly shown. The obtained temperature is the half of
the accepted value of the Hawking temperature. It is demonstrated that how
the proper regularization of singular integrals resolves the discrepancy in
the aforementioned temperatures. Sec. IV and V are devoted to the
calculation of the Hawking temperature in PG and IEF coordinate systems,
respectively. In Sec. VI we apply the HJ method to KS form of the LDBHs.
Finally, the conclusion and future directions are given in Sec. VII.

\section{\textbf{LDBH and HJ Method}}

In general, the metric of spherically symmetric and static BH in four
dimensions is given by

\begin{equation}
ds^{2}=-fdt^{2}+f^{-1}dr^{2}+R^{2}d\Omega ^{2},  \label{1n}
\end{equation}

where 
\begin{equation}
d\Omega ^{2}=d\theta ^{2}+\sin ^{2}\theta d\varphi ^{2},  \label{2n}
\end{equation}

is the metric on a unit two-sphere $S^{2}$. Since we aim to solve the
relativistic HJ equation for a massive but uncharged scalar field in the
LDBH background, let us first analyze the geometry of the LDBH. Whenever the
metric functions of the line-element (1) are given by

\begin{equation}
R^{2}=A^{2}r,\text{ \ \ \ \ }f=\Sigma (r-r_{h}),  \label{3n}
\end{equation}

we call the metric (1) as LDBH \cite{Clement,MSH}. In various theories (EMD,
EYMD and EYMBID), metric functions do not alter their forms as seen in the
Eq. (3). Only non-zero positive constants $A$ and $\Sigma $ take different
values depending on which theory is taken into account \cite{MSH}. It can be
easily deduced from the metric function $f$ \ that LDBHs possess NAF
geometry and $r_{h}$ represents the horizon. For $r_{h}\neq 0$, the horizon
hides the null singularity at $r=0$. Even in the extreme case $r_{h}=0$ in
which the central null singularity at $r=0$ is marginally trapped, such that
outgoing waves are permitted to reach the external observers, the LDBH still
sustains its BH property.

NAF structure of the LDBH leads us to use the definition of quasi-local mass 
$M$ \cite{BrownYork} in order to obtain a relationship between the horizon $%
r_{h}$\ and the mass $M$ of the BH as follows%
\begin{equation}
r_{h}=\frac{4M}{\Sigma A^{2}},  \label{4n}
\end{equation}

According to the laws of BH thermodynamics, the conventional definition of
the Hawking temperature $T_{H}$ \cite{Wald} is expressed in terms of the
surface gravity $\kappa $. For the metric (1), $T_{H}$ is given in the
explicit form as follows:

\begin{equation}
T_{H}=\frac{\kappa }{2\pi }=\left. \frac{\partial _{r}f}{4\pi }\right\vert
_{r=r_{h}},  \label{5n}
\end{equation}

After substituting the metric function $f$ (3) into the above equation, $%
T_{H}$ of the LDBH becomes

\begin{equation}
T_{H}=\frac{\Sigma }{4\pi }.  \label{6n}
\end{equation}

We can immediately observe that the obtained temperature is constant. In
general, such an event typically occurs in an isothermal process of the
standard thermodynamics in which $\Delta T=0$. Therefore, the LDBH's
radiation is such a particular process that the energy (mass) transfer out
of the BH typically happens at a slow rate that thermal equilibrium is
maintained.

Here, we consider the problem of a scalar particle which it moves in this
spacetime while there is no back-reaction or self-gravitational effect.
Within the semi-classical framework, the classical action $I$ of the
particle satisfies the relativistic HJ equation\ 

\begin{equation}
g^{\mu \nu }\partial _{\mu }I\partial _{\nu }I+m^{2}=0,  \label{7n}
\end{equation}

in which $m$ is the mass of the scalar particle, and $g^{\mu \nu }$
represents the invert metric tensors derived from the metric (1). By
considering Eqs.(1), (3) and (7), we get%
\begin{equation}
\frac{-1}{f}(\partial _{t}I)^{2}+f(\partial _{r}I)^{2}+\frac{1}{A^{2}r}\left[
(\partial _{\theta }I)^{2}+\frac{1}{\sin ^{2}\theta }(\partial _{\varphi
}I)^{2}\right] +m^{2}=0,  \label{8n}
\end{equation}

For the HJ equation it is common to use the separation of variables method
for the action $I=I(t,r,\theta ,\varphi )$ as follows

\begin{equation}
I=-Et+W(r)+J(x^{i}),  \label{9n}
\end{equation}

where

\begin{equation}
\partial _{t}I=-E,\text{ \ \ \ \ \ }\partial _{r}I=\partial _{r}W(r),\text{
\ \ \ \ \ }\partial _{i}I=J_{i},  \label{10n}
\end{equation}

and $J_{i}$'s are constants in which $i=1,2$ labels angular coordinates $%
\theta $ and $\varphi $, respectively. Since the norm of the timelike
Killing vector $\partial _{t}$ is (negative) unity at a particular location $%
r\equiv \check{r}=\frac{1}{\Sigma }+r_{h}$, $E$\ is the energy of the
particle detected by an observer at $\check{r}$, where is outside the
horizon. Solving for $W(r)$ yields

\begin{equation}
W(r)=\pm \int \frac{\sqrt{E^{2}-\frac{f}{A^{2}r}[J_{\theta }^{2}+\frac{%
J_{\varphi }^{2}}{\sin ^{2}\theta }+\left( mA\right) ^{2}r]}}{f}dr,
\label{11n}
\end{equation}

where $\pm $ naturally comes since the Eq.(8) was quadratic in terms of $%
W(r) $. Solution of the Eq.(11) with "$+$" sign corresponds to scalar
particles moving away from the BH (outgoing) and the other solution i.e.,
the solution with "$-$" sign represents particles which move toward the BH
(ingoing). After evaluating the above integral around the pole at the
horizon (adhering to the Feynman's prescription \cite{Feynman}), one arrives
at the following:

\begin{equation}
W_{\left( \pm \right) }=\pm \frac{i\pi E}{\Sigma }+c,  \label{12n}
\end{equation}

where $c$ is a complex integration constant. Thus, we can deduce that
imaginary parts of the action can arise due to the pole at the horizon and
from the complex constant $c$. Thence, we can determine the probabilities of
ingoing and outgoing particles while crossing the horizon as

\begin{equation}
P_{out}=e^{-2\func{Im}I}=\exp \left[ -2(\func{Im}W_{\left( +\right) }+\func{%
Im}c)\right] ,  \label{13n}
\end{equation}

\begin{equation}
P_{in}=e^{-2\func{Im}I}=\exp \left[ -2(\func{Im}W_{\left( -\right) }+\func{Im%
}c)\right] ,  \label{14n}
\end{equation}

In the classical point of view, a BH absorbs any ingoing particles passing
its horizon. In other words, there is no reflection for the ingoing waves
which corresponds to $P_{in}=1$. This is enabled by setting $\func{Im}c=%
\frac{\pi E}{\Sigma }.$ This choice also implies that the imaginary part of
the action $I$ for a tunneling particle can only come out $W_{+}$. Namely,
we get 
\begin{equation}
\func{Im}I=\func{Im}W_{+}=\frac{2\pi E}{\Sigma },  \label{15n}
\end{equation}

which is independent of the horizon $r_{h}$. Therefore, the tunneling rate
for the LDBH can be obtained as

\begin{equation}
\Gamma =P_{out}=e^{\frac{-4\pi E}{\Sigma }},  \label{16n}
\end{equation}

and since \cite{PW}

\begin{equation}
\Gamma =e^{-\beta E},  \label{17n}
\end{equation}

in which $\beta $ denotes the Boltzmann factor and $T=\frac{1}{\beta }$, one
can easily read the horizon temperature of the LDBH as 
\begin{equation}
\check{T}_{H}=\frac{\Sigma }{4\pi }.  \label{18n}
\end{equation}

which means that Hawking temperature $T_{H}$ (6) is impeccably recovered.

\section{Isotropic Coordinates}

In general, when the metric (1) is transformed to the isotropic coordinates,
the resulting line-element admits a BH spacetime in which the metric
functions are non-singular at the horizon, the time direction is a Killing
vector and the three dimensional subspace of the spatial part of the
line-element (known as time slice) appears as Euclidean with a conformal
factor. Furthermore, using of these coordinates renders the calculation of
the index of refraction of the light rays (a subject of gravitational
lensing) around a BH possible. So, the light propagation of a BH can be
mimicked by the index of refraction. By this way, an observer may identify
the type of the BH \cite{Perlick}.

In this section, we firstly transform the LDBH to the isotropic coordinates
and then analyze the HJ equation. Next, we examine the horizon temperature
whether it agrees with the $T_{H}$ or not. At the final stage, we discuss
the discrepancy in the temperatures and its abolishment.

The LDBH solution in isotropic coordinates can be found by the following
transformation

\begin{equation}
\frac{d\rho }{\rho }=\frac{dr}{A\sqrt{\Sigma (r^{2}-rr_{h})}},  \label{19n}
\end{equation}

so that we obtain

\begin{equation}
\rho =\left( 2r-r_{h}+2\sqrt{r(r-r_{h})}\right) ^{\frac{1}{\gamma }},
\label{20n}
\end{equation}

and inversely

\begin{equation}
r=\frac{1}{4\rho ^{\gamma }}\left( \rho ^{\gamma }+r_{h}\right) ^{2},
\label{21n}
\end{equation}

where $\gamma =A\sqrt{\Sigma }$. This transformation takes the metric (1) to
the form

\begin{equation}
ds^{2}=-Fdt^{2}+G(d\rho ^{2}+\rho ^{2}d\Omega ^{2}),  \label{22n}
\end{equation}

with

\begin{equation}
F=\frac{\Sigma }{4\rho ^{\gamma }}\left( \rho ^{\gamma }-r_{h}\right) ^{2},%
\text{ \ \ \ \ \ }G=\frac{A^{2}}{4\rho ^{\gamma +2}}\left( \rho ^{\gamma
}+r_{h}\right) ^{2},  \label{23n}
\end{equation}

In this coordinate system, the event horizon is located at $\rho _{h}=\left(
r_{h}\right) ^{\frac{1}{\gamma }}$ and the region $\rho >\rho _{h}$ covers
the exterior region of the LDBH, which is static. In the naive coordinates
(1) of the LDBH, all Killing vectors are spacelike in the interior region
and we deduce that the interior of the LDBH is nonstationary. On the other
hand, when we consider the interior region $\rho <\rho _{h}$ of the metric
(22), it admits a hypersurface-orthogonal timelike Killing vector which
implies the static region. Namely, the region $\rho <\rho _{h}$ does not
cover the interior of the LDBH. Instead, it again covers the exterior region
such that metric (22) is a double covering of the LDBH exterior \cite{Edery}%
.\ 

One can easily rewrite the metric (22) as follows

\begin{equation}
ds^{2}=F(-dt^{2}+\widehat{g}),  \label{24n}
\end{equation}

and obtain the Fermat metric \cite{Perlick}\ form of the LDBH as

\begin{equation}
\widehat{g}=n(\rho )^{2}(d\rho ^{2}+\rho ^{2}d\Omega ^{2}),  \label{25n}
\end{equation}

where $n(\rho )$ is known as the index of refraction. For the LDBH medium,
it is calculated as

\begin{equation}
n(\rho )=\sqrt{\frac{G}{F}}=\frac{A}{\sqrt{\Sigma }\rho }\frac{\rho ^{\gamma
}+r_{h}}{\rho ^{\gamma }-r_{h}},  \label{26n}
\end{equation}

The expression for the HJ equation (7) on the background (22) corresponds to

\begin{equation}
\frac{-1}{F}(\partial _{t}I)^{2}+\frac{1}{G}(\partial _{\rho }I)^{2}+\frac{1%
}{G\rho ^{2}}\left[ (\partial _{\theta }I)^{2}+\frac{1}{\sin ^{2}\theta }%
(\partial _{\varphi }I)^{2}\right] +m^{2}=0,  \label{27n}
\end{equation}

There exists a solution of the form

\begin{equation}
I=-Et+W_{iso}(\rho )+J(x^{i}),  \label{28n}
\end{equation}

Solving for $W_{iso}(\rho )$\ yields

\begin{equation}
W_{iso}(\rho )=\pm \dint n(\rho )\sqrt{E^{2}-\frac{F}{G\rho ^{2}}\left(
J_{\theta }^{2}+\frac{J_{\varphi }^{2}}{\sin ^{2}\theta }\right) -m^{2}F}%
d\rho ,  \label{29n}
\end{equation}

which can be written near the horizon $\rho \approx \left( r_{h}\right) ^{%
\frac{1}{\gamma }}$ as

\begin{equation}
W_{iso(\pm )}=\pm E\int n(\rho )d\rho ,  \label{30n}
\end{equation}

Here, it is clear that $W_{iso(\pm )}$ is governed by the index of
refraction of the LDBH. After applying the Feynman's prescription to the
above integral, one obtains

\begin{equation}
W_{iso(\pm )}=\pm \frac{i2\pi E}{\Sigma }+c_{2},  \label{31n}
\end{equation}

where $c_{2}$ is another integration constant. Similar to the procedure
followed in the previous section i.e., setting $P_{in}=1$ which yields $%
\func{Im}c_{2}=\frac{2\pi E}{\Sigma }$, we obtain the imaginary part of the
action $I$\ of the tunneling particle as follows

\begin{equation}
\func{Im}I=\func{Im}W_{iso(+)}=\frac{4\pi E}{\Sigma },  \label{32n}
\end{equation}

Thus, by using the tunneling rate formulation (16) one obtains the horizon
temperature of the LDBH as

\begin{equation}
\check{T}_{H}=\frac{\Sigma }{8\pi },  \label{33n}
\end{equation}

But the obtained temperature is the half of the conventional Hawking
temperature, $\check{T}_{H}=\frac{1}{2}T_{H}$. So, the above result (33)
represents that transforming the naive coordinates to the isotropic
coordinates yield an apparent temperature of the BH that it is less than the
true temperature $T_{H}$. This is in analogy with the apparent depth $q$ of
a fish swimming at a depth $d$ below the surface of a pond is less than the
true depth $d$ \ i.e., $q<d$. This illusion is due to the difference of the
index of refractions between the mediums. Particularly, such a case occurs
when $n_{observer}<n_{object},$ as in here. Because, it is obvious from
Eq.(26) that the index of refraction of the medium of an observer who is
located at the outer region is less than the index of refraction of the
medium near to the horizon. Since the value of $W_{iso(\pm )}$ (30) acts as
a decision-maker on the value of the Hawking temperature $T_{H}$\ of the BH,
one can deduce that the index of refraction (26), consequently the
gravitational lensing effect, plays an important role on the observation of
the true $T_{H}$.

On the other hand, we admittedly know that coordinate transformation of the
naive coordinates to the isotropic coordinates should not alter the true
temperature of the BH. Since the appearances are deceptive, one should make
deeper analysis to get the real. Very recently, this problem has been
throughly discussed by Chatterjee and Mitra \cite{GRG12}. Since the
isotropic coordinate $\rho $ becomes complex inside the horizon ($r<r_{h}$),
they have proven that while evaluating the integral (30) around the horizon,
the path across the horizon involves a change of $\pi /2$ instead of $\pi $
in the phase of the complex variable $\left( \rho ^{\gamma }-r_{h}\right) $.
This could best be seen from Eq.(21), which is rewritten as

\begin{equation}
r=r_{h}+\frac{(\rho ^{\gamma }-r_{h})^{2}}{4\rho ^{\gamma }},  \label{34n}
\end{equation}

and implies that

\begin{eqnarray}
\frac{dr}{r-r_{h}} &=&-\gamma \frac{d\rho }{\rho }+\frac{2\gamma d\rho }{%
\rho ^{1-\gamma }(\rho ^{\gamma }-r_{h})},  \notag \\
&=&-\gamma \frac{d\rho }{\rho }+\frac{2dz}{z-r_{h}}.  \label{35n}
\end{eqnarray}

where $z=\rho ^{\gamma }$. The first term does not admit any imaginary part
at the horizon. Hence, any imaginary contribution coming from $\frac{2dz}{%
z-r_{h}}$ must be half of the $\frac{dr}{r-r_{h}}$. The latter remark
produces a factor $i\pi /2$ for the integral (30) and subsequently it yields 
$\func{Im}W_{iso(+)}=\frac{2\pi E}{\Sigma }$ as obtained in the previous
section. Thus, we get the horizon temperature as $\check{T}_{H}=\frac{\Sigma 
}{4\pi }$ which is again the $T_{H}$.

\section{\textbf{PG Coordinates }}

Generally, we use the PG coordinates \cite{Painleve, Gullstrand} in order to
describe the spacetime on either side of the event horizon of a static BH.
In this coordinate system, an observer does not consider the surface of the
horizon to be in any way special. In this section, we shall employ the PG
coordinates as another regular coordinate system in the HJ equation and
examine whether they yield the correct calculation of the $T_{H}$\ or not.

We can pass to the PG coordinates by applying the following transformation 
\cite{Robertson} to the metric (1)

\begin{equation}
dT=dt+\frac{\sqrt{1-f}}{f}dr,  \label{36n}
\end{equation}

where $T$ is our new time coordinate (let us say PG time). Substituting this
metric (1) gives 
\begin{equation}
ds^{2}=-fdT^{2}+2\sqrt{1-f}dTdr+dr^{2}+R^{2}d\Omega ^{2},  \label{37n}
\end{equation}

One of the main features of these coordinates is that the PG time
concurrently corresponds to the proper time. For the metric (37) the HJ
equation (7) takes the form

\begin{equation}
-(\partial _{T}I)^{2}+2\sqrt{1-f}(\partial _{T}I)(\partial _{r}I)+f(\partial
_{r}I)^{2}+\frac{1}{R^{2}}(\partial _{\theta }I)^{2}+\frac{1}{R^{2}\sin
^{2}\theta }(\partial _{\varphi }I)^{2}+m^{2}=0,  \label{38n}
\end{equation}

Letting

\begin{equation}
I=-ET+W_{PG}(r)+J(x^{i}),  \label{39n}
\end{equation}

and substitute Eqs.(39) and (3) into Eq.(38), we find

\begin{equation}
W_{PG}(r)=\int \frac{E}{\Sigma (r-r_{h})}\left( \sqrt{1-\Sigma (r-r_{h})}\pm 
\sqrt{1-\Sigma (r-r_{h)}-\frac{\lambda \Sigma (r-r_{h)}}{E^{2}}}\right) dr,
\label{40n}
\end{equation}

where

\begin{equation}
\lambda =m^{2}-E^{2}+\frac{J_{\theta }^{2}}{R^{2}}+\frac{J_{\varphi }^{2}}{%
R^{2}\sin ^{2}\theta },  \label{41n}
\end{equation}

Near the horizon, Eq.(40) in turn implies that

\begin{equation}
W_{PG(\pm )}=\frac{E}{\Sigma }\int \frac{1}{(r-r_{h})}(1\pm 1)dr,
\label{42n}
\end{equation}

Therefore, imposing the condition $W_{PG(-)}=0$ which ensures that there is
no reflection for the ingoing particle, we have

\begin{equation}
W_{PG(+)}=\frac{i2\pi E}{\Sigma }.  \label{43n}
\end{equation}

Thus, we get the imaginary part of the action $I$ as

\begin{equation}
\func{Im}I=\func{Im}W_{PG(+)}=\frac{2\pi E}{\Sigma },  \label{44n}
\end{equation}

With the aid of Eqs.(16) and (17), one can readily read the horizon
temperature of the LDBH which is featured in the PG\ coordinates as

\begin{equation}
\check{T}_{H}=\frac{\Sigma }{4\pi }.  \label{45n}
\end{equation}

This result is fully in agreement with the standard value of the Hawking
temperature (6).

\section{IEF Coordinates}

The another useful coordinate system which is also regular at the event
horizon originally constructed by Eddington \cite{Eddington} and Finkelstein 
\cite{Finkelstein}. These coordinates are fixed to radially moving photons.
The line-element (1) takes the following form in the IEF coordinates (see
for instance \cite{Poisson})

\begin{equation}
ds^{2}=-fd\upsilon ^{2}+2\sqrt{1-f}d\upsilon dr+dr^{2}+R^{2}(d\theta
^{2}+\sin ^{2}\theta d\varphi ^{2}),  \label{46n}
\end{equation}

in which $\upsilon $ is a new null coordinate, the so-called advanced time.
It is given by

\begin{equation}
\upsilon =t+r_{\ast },  \label{47n}
\end{equation}

where $r_{\ast }$ is known as the Regger-Wheeler coordinate or the tortoise
coordinate. For the outer region of the LDBH, it is found to be

\begin{equation}
r_{\ast }=\frac{1}{\Sigma }\ln \left( \frac{r}{r_{h}}-1\right)  \label{48n}
\end{equation}

Since the metric (46) has a Killing vector field of $\xi ^{\mu }=\partial
_{\upsilon }$, in this coordinate system an observer measures the scalar
particle's energy as $E=-\partial _{\upsilon }I$. In this regard, the action
is assumed to be of the form

\begin{equation}
I=-E\upsilon +W_{EF}(r)+J(x^{i}).  \label{49n}
\end{equation}

Employing the HJ equation (7) in the metric (46), the final result for $%
W_{EF}(r)$ can be found as

\begin{equation}
W_{EF}(r)=\int \frac{E}{\Sigma (r-r_{h})}\left( 1\pm \sqrt{1-\frac{\varkappa
\Sigma (r-r_{h)}}{E^{2}}}\right) dr  \label{50n}
\end{equation}

where

\begin{equation}
\varkappa =m^{2}+\frac{J_{\theta }^{2}}{R^{2}}+\frac{J_{\varphi }^{2}}{%
R^{2}\sin ^{2}\theta },  \label{51n}
\end{equation}

In the vicinity of the event horizon, $W_{EF}(r)$ reduces to the following
expression

\begin{equation}
W_{EF(\pm )}=\frac{E}{\Sigma }\int \frac{1}{(r-r_{h})}(1\pm 1)dr,
\label{52n}
\end{equation}

which has the same expression appeared in the Eq. (42). Henceforth,

\begin{equation}
W_{EF(-)}=0\text{, \ \ }W_{EF(+)}=\frac{i2\pi E}{\Sigma }\text{ \ }%
\rightarrow \text{\ \ }\func{Im}I=\func{Im}W_{EF(+)}=\frac{2\pi E}{\Sigma },
\label{53n}
\end{equation}

and likewise to the PG coordinates, the usage of the EF coordinates in the
HJ equation enables us to reproduce the standard Hawking temperature from
the horizon temperature of the LDBH:

\begin{equation}
\check{T}_{H}=\frac{\Sigma }{4\pi }=T_{H}.  \label{54n}
\end{equation}

\section{KS Coordinates}

The another well-behaved coordinate system which covers the entire spacetime
manifold of the maximally extended BH solution is the so-called KS
coordinates \cite{Kruskal, Szekeres}. These coordinates have the effect of
squeezing infinity into a finite distance and thus the entire spacetime can
be displayed on a stamp-like diagram. In this section, we will apply the HJ
equation to KS metric of the LDBH in order to verify whether $\check{T}_{H}$
is going to be equal to the $T_{H}$ or not.

Metric (1) can be rewritten as follows \cite{Poisson}

\begin{equation}
ds^{2}=-fdudv+R^{2}d\Omega ^{2},  \label{55n}
\end{equation}

where

\begin{equation}
du=dt-dr_{\ast }\text{, \ \ \ \ }dv=dt+dr_{\ast },  \label{56n}
\end{equation}

Furthermore, if we define new coordinates $(U,V)$ in terms of the surface
gravity $\kappa $\ (5)

\begin{equation}
U=-e^{-\kappa u}\text{, \ \ \ \ }V=e^{\kappa v},  \label{57n}
\end{equation}

metric (55) transforms to the KS metric as

\begin{equation}
ds^{2}=\frac{f}{\kappa ^{2}}\frac{dUdV}{UV}+R^{2}d\Omega ^{2},  \label{58n}
\end{equation}

Recalling the definitions given in Eqs. (3-5) and (57), it is then
straightforward to obtain the KS metric of the LDBH. It is given by

\begin{equation}
ds^{2}=-\frac{16M}{\Sigma ^{2}A^{2}}dUdV+R^{2}d\Omega ^{2},  \label{59n}
\end{equation}

This metric is well-behaved everywhere outside the physical singularity $r=0$%
. Alternatively, metric (59) can be rewritten as

\begin{equation}
ds^{2}=-dT^{2}+dX^{2}+R^{2}d\Omega ^{2},  \label{60n}
\end{equation}

This is possible with the following transformation

\begin{equation}
T=\frac{4\sqrt{M}}{\Sigma A}(V+U)=\frac{4\sqrt{M}}{\Sigma A}\sqrt{\frac{r}{%
r_{h}}-1}\sinh (\frac{\Sigma t}{2}),  \label{61n}
\end{equation}

\begin{equation}
X=\frac{4\sqrt{M}}{\Sigma A}(V-U)=\frac{4\sqrt{M}}{\Sigma A}\sqrt{\frac{r}{%
r_{h}}-1}\cosh (\frac{\Sigma t}{2}),  \label{62n}
\end{equation}

From now on, it easy to see that

\begin{equation}
X^{2}-T^{2}=\frac{16M}{\Sigma ^{2}A^{2}}(\frac{r}{r_{h}}-1),  \label{63n}
\end{equation}

which means that $X=\pm T$ corresponds to the future and past horizons. On
the other hand, $\partial _{T}$\ is not a timelike Killing vector anymore
for the metric (60), instead one should consider the timelike Killing vector
as

\begin{equation}
\partial _{\widehat{T}}=N(X\partial _{T}+T\partial _{X}),  \label{64n}
\end{equation}

where $N$ denotes the normalization constant. It can admit a specific value
that the norm of the Killing vector becomes negative unity at a specific
location in the outer region of the LDBH where $r=\frac{1}{\Sigma }+r_{h}$.
This implies that

\begin{equation}
N=\frac{\Sigma }{2},  \label{65n}
\end{equation}

Since the energy is defined by 
\begin{equation}
\partial _{\widehat{T}}I=-E,  \label{66n}
\end{equation}

then

\begin{equation}
\frac{\Sigma }{2}(X\partial _{T}I+T\partial _{X}I)=-E,  \label{67n}
\end{equation}

Without loss of generality, we may only consider the (1+1) dimensional form
of the KS metric (60) which now appears as Minkowskian

\begin{equation}
ds^{2}=-dT^{2}+dX^{2},  \label{68n}
\end{equation}%
The calculation of the HJ method is more straightforward in this case. The
HJ equation (7) for the above metric reads

\begin{equation}
-\left( \partial _{T}I\right) ^{2}+\left( \partial _{X}I\right) ^{2}+m^{2}=0,
\label{69n}
\end{equation}

This equation implies that the action $I$ to be used in the HJ equation (7)
for the metric (68) can be

\begin{equation}
I=g(X-T)+J(x^{i}),  \label{70n}
\end{equation}

For simplicity, we may further set $J(x^{i})=0$ and $m=0.$ Using Eq. (67)
with the ansatz (70), one derives the following expression.

\begin{equation}
g(u)=\int \frac{2E}{\Sigma u}du,  \label{71n}
\end{equation}

where $u=X-T$. This expression develops a divergence at the horizon $u=0$,
namely $X=T$. Thus, it leads to a pole at the horizon (doing a semi-circular
contour of integration in the complex plane) and the result is found to be

\begin{equation}
\func{Im}I=\frac{2\pi E}{\Sigma },  \label{72n}
\end{equation}

So, referring to the tunneling probability (7) we get

\begin{equation}
\Gamma =e^{\frac{-4\pi E}{\Sigma }}.  \label{73n}
\end{equation}

which means that the correct Hawking temperature $T_{H}=\frac{\Sigma }{4\pi }
$ is recovered in the background of the KS metric of the LDBH.

\section{Conclusion}

In this study, the Hawking radiation of the LDBH in four dimensions via the
HJ method is studied. To the authors's knowledge, LDBH is the only BH that
its radiation obeys an isothermal process which corresponds to no change in
the temperature during its evaporation: $\Delta T=0$. This can be easily
deduced from its Hawking temperature which yields constant value. Namely, it
is independent from the mass $M$ (or horizon $r_{+}$) of the BH. In addition
to the naive coordinates, four different regular coordinate systems are
examined throughout this study. It was shown that the computed horizon
temperatures in the naive, PG, IEF and KS coordinates via the HJ method
exactly matched with the conventional Hawking temperature. Here, we should
notice that in the Sec. VI which considers the KS coordinates, the way that
followed up was slightly different than the other sections. In that section,
without loss of generality, we discarded the mass of the scalar field and
neglected the angular dependence of the HJ equation. This turned out to be
the application of the HJ method for the Minkowski metric. As a result,
matching of the temperatures was successfully shown.

We believe that the most interesting part of the present paper is Sec. III
where the LDBH metric was expressed in terms of the isotropic coordinates.
Using of the Fermat metric enabled us to determine the index of refraction
of the LDBH. In particular, it is proven that the index of the refraction
plays a decisive role on the tunneling rate. Unlike to the other coordinate
systems, in the isotropic coordinates the standard integration around the
pole at the horizon caused to produce unacceptable value of the temperature:
half of the standard $T_{H}$. In order to overcome this discrepancy, we
inspired from a recent study \cite{GRG12} which has demonstrated how the
proper regularization of singular integrals leads to the standard Hawking
temperature for the isotropic coordinates. As a result, it is clarified that
the path across the horizon entails the value $\frac{i\pi }{2}$ on
integration instead of $i\pi $. The underlying reason of this is that the
isotropic radial coordinate $\rho $ (20)\ is real outside the BH, however it
becomes complex inside the BH.

Finally, it would be interesting to extend our analysis to yet another BHs,
which could be BHs with multiple horizons, multi-BHs, higher dimensional BHs
etc. This will be considered in the near future.

\end{document}